\documentclass{aa501}
\usepackage{graphicx}

\usepackage{natbib}
\bibpunct[, ]{(}{)}{;}{a}{}{,}

\usepackage[german,english]{babel}
\newcommand{\gppr}{\stackrel{>}{\scriptstyle \sim}}
\newcommand{\gappr}{\raisebox{-0.4ex}{$\gppr$}}

\newcommand{\Mwd}{\mbox{$M_\mathrm{wd}$}}

\begin{document}

\title{Implications of the HST/FGS parallax
of SS\,Cygni\\
on the disc instability model}

\author{M.R. Schreiber \inst{1,2} \and B.T. G\"ansicke \inst{1}}

\offprints{M.R. Schreiber, email: mschrei@uni-sw.gwdg.de}

\institute{Universit\"ats-Sternwarte, Geismarlandstr.11, D-37083 G\"ottingen,
Germany \and Department of Astronomy, University of Cape Town, Private Bag,
Rondebosch 7700, Cape Town, South Africa}

\date{Received \underline{\hskip2cm} ; accepted \underline{\hskip2cm} }

\titlerunning{The HST/FGS parallax of SS\,Cygni and the DIM}

\authorrunning{Schreiber \& G\"ansicke}

\abstract{ 
We analyse the consequences of the recently measured
parallax of SS\,Cygni \citep{harrisonetal99-1} on the accretion disc
limit cycle model.  Using the observed long term light curve of
SS\,Cyg and $d=166\pm\,12\mathrm{pc}$, we obtain for the mean mass
transfer rate
%BG%,
$\overline{\dot{M}}_{\mathrm{tr}} =
4.2\pm\,1.7\times\,10^{17}\mathrm{g\,s^{-1}}$.
In addition, we calculate the vertical structure of the accretion disc
taking into account heating of the outer disc by the stream impact.
Comparing the mean accretion rate derived from the observations with
the calculated critical mass transfer rate, we
find that the disc instability model disagrees with the observed long
term light curve of SS\,Cyg as
$\overline{\dot{M}}_{\mathrm{tr}}$ is greater or similar to the critical mass
transfer rate. 
The failure of the model indicated by this result can be confirmed by
considering that the accretion rate at the onset of the decline should
be exactly equal to the value critical for stability. In contrast to
this prediction of the model, we find that the accretion rate required
to explain the observed visual magnitude at the onset of the decline {\em
must} be significantly higher than the critical mass transfer rate.
Our results strongly suggest that either the usually assumed temperature
dependence of the viscosity parameter $\alpha$ is not a realistic description
of the disc viscosity,  
that the mass transfer rate in SS\,Cyg noticeably
increases during the outbursts or, finally, that the HST distance of
$166\pm\,12$\,pc, is too high. 
}

\maketitle

\section{Introduction}
In non-magnetic {\em cataclysmic variables} (CVs) \citep[][for an
encyclopaedic review]{warner95-1} a white dwarf accretes material from
the Roche-lobe filling secondary via an accretion disc. In many CVs,
sudden brightenings -- called dwarf nova outbursts -- are observed.
These eruptions are generally thought to result from thermal
instabilities in the accretion discs \citep[the disc instability model
(DIM), see e.g.][for parameter studies]{cannizzo93-1,ludwigetal94-1}.
Calculations of the steady-state solution of the vertical disc
structure for a given radius show a hysteresis relation in the
surface-density vs. accretion rate plane
%diagram 
-- the so-called ``S-curve''.  In principle, a limit cycle behaviour
can be maintained by a mean mass-transfer rate which is between the
critical rate needed to stay in a cold quiescent state and that needed
to remain in the hot outburst state.  However, the obtained light
curves using a constant viscosity parameter $\alpha$ show only low
amplitude outbursts \citep{smak84-2}.  In order to match the observed
outburst light curves, it is necessary to assume a drastic change of
the viscosity parameter $\alpha$ between the hot and the cold state,
$\alpha_\mathrm{hot}$ and $\alpha_\mathrm{cold}$, with
$\alpha_\mathrm{hot}/\alpha_\mathrm{cold}$ typically 5--10 but with unexplained exceptions
$\alpha_\mathrm{hot}/\alpha_\mathrm{cold}\sim\,10^{3}$ e.g. for
WZ\,Sge.  Although the change of $\alpha$ is a somewhat arbitrary
assumption, the DIM has become the generally accepted explanation
of dwarf nova outbursts as it agrees with observational and
theoretical constraints \citep[see][]{cannizzo93-1}.
Recently, additional effects such as disc radius variations
\citep{hameuryetal98-1}, irradiation
\citep{hameuryetal99-1,schreiberetal00-2,schreiber+gaensicke01-1},
mass transfer variations \citep{schreiberetal00-1,buat-menardetal01-2}
and the effects of the accretion stream impact
\citep{schreiber+hessman98-1,buat-menardetal01-1,stehleetal01-1} have been
included in the model. 

\section{The absolute brightest dwarf nova: SS\,Cyg}
The dwarf nova SS\,Cyg is among the visually brightest CVs and has a
detailed long term light curve covering more than a century. SS\,Cyg
has, therefore, become \textit{the} reference object for accretion
disc instability analyses, and its observed light curve has been used
to calibrate the viscosity parameter $\alpha$.
\citet{cannizzo+mattei92-1} examined the long term light curve of
SS\,Cyg (Fig.\,1) in detail in order to test the limit
cycle model \citep{cannizzo93-2} and found a mean outburst duration
of $t_{\mathrm{out}}=10.76$\,d and a mean quiescence time of
$t_{\mathrm{qui}}=38.71$\,d (corresponding to a mean cycle duration of
$t_{\mathrm{cyc}}=49.47$\,d). In a later paper,
\citet{cannizzo+mattei98-1} again examined the light curve of
SS\,Cyg, deriving the rise and decay times. Only the mean values are
important in the context of our analysis: $t_{\mathrm{ris}}=0.5$\,d,
$t_{\mathrm{dec}}=2.5$\,d.  These values are obtained using a
conservatively low estimate of the outburst magnitude of $m_v=8.5$. 

\begin{figure}
\label{f-longterm}
\includegraphics[width=8.8cm]{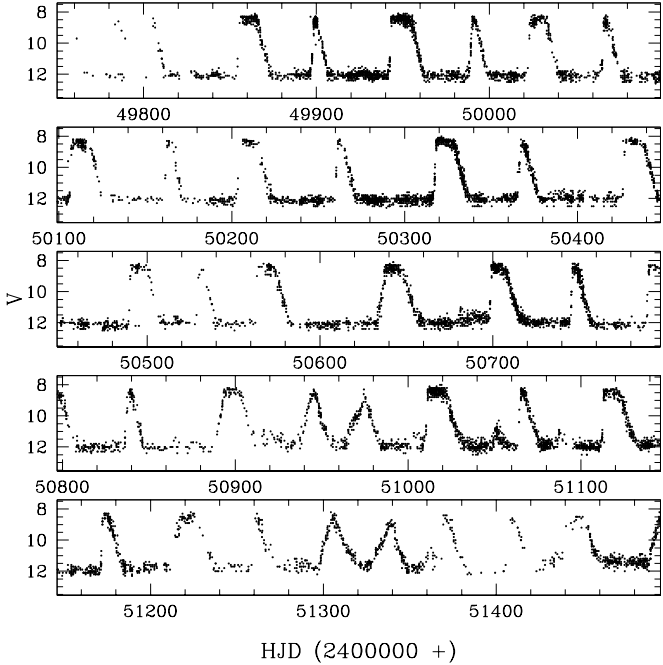}
\caption[]{The visual long term light curve of SS\,Cyg compiled from
observations made by members of the Association Francaise des
Observateurs d'Etoiles Variables.}
\end{figure}

However, even though the brightness variation of SS\,Cyg has been
monitored to a high level of precision, its \textit{distance} has been
highly uncertain. As a consequence, the absolute
magnitude and accretion rates, 
have also been ill-defined. Recently, a high-quality parallax of
SS\,Cyg was obtained using the Hubble Space Telescope (HST) Fine
Guidance Sensor (FGS), which moves the system to a distance $1.5-2$
times larger than previously thought \citep{harrisonetal99-1}. The
HST/FGS distance of $166\pm12$\,pc is confirmed by
\citet{beuermann00-1}, who showed that the accretion disc in SS\,Cyg
contributes significantly to the observed $K$ band flux.
As a consequence of the now firmly established (larger) distance, the
absolute magnitude of SS\,Cyg in outburst increased, too.
Figure\,2 shows the absolute visual magnitudes of dwarf
novae at outburst maximum as a function of their orbital period,
adapted from \citet{warner87-1}.  The arrow displays the impact
of the HST/FGS parallax on SS\,Cyg: the system is now the
absolute brightest dwarf nova in outburst and its maximum absolute magnitude
is brighter than one would expect from its orbital period
\citep{cannizzo98-1}.    

The mean accretion rate in SS\,Cyg inferred from its long term
light curve depends, in addition to the absolute magnitude, on the
inclination $i$ of the accretion disc. This system parameter is
somewhat uncertain, most studies of SS\,Cyg suggest
$i=37^{\circ}\pm\,5^{\circ}$ \citep{ritter+kolb98-1}. A significantly
higher value of $i\simeq57^{\circ}$ was derived by
\citet{voloshina+khruzina97-1}, but, throughout this paper we assume
the lower value.  To account for the influence of the inclination on
the absolute magnitude we follow
\citet{paczynski+schwarzenberg-czerny80-1}:
\begin{equation}
\Delta\,M_{\mathrm{v}}(i)=-2.5\log\left[\frac{(0.4+0.6\cos\,i)\cos\,i}{0.4}\right],
\end{equation} 
where we used a limb darkening coefficient $u=0.6$. Notice, Eq.\,(1) gives
the inclination correction corresponding to a mean inclination
$i=56^{\circ}.7$ (Fig.\,2).  

Throughout this paper we use the system parameters of SS\,Cyg
given in the literature, i.e.
the primary mass $M_{\mathrm{wd}}=1.19\pm\,0.05\,M_{\odot}$ 
\citep{ritter+kolb98-1,vanteeseling97-1},
secondary mass $M_{\mathrm{sec}}=0.70\,\,M_{\odot}$,
and orbital period in hours 
$P_{\mathrm{hr}}=6.6$ \citep{ritter+kolb98-1}.

\section{Deriving the critical and the mean mass transfer rate}
The goal of this section is to check if the ``standard'' disc
instability model can reproduce the absolute magnitude of the
outbursts of SS\,Cyg, taking into account the larger HST/FGS
distance, the inclination, and the mean outburst time scales derived
from the long term light curve.

\begin{figure}
\label{f-mvorb}
\includegraphics[angle=270,width=8.65cm]{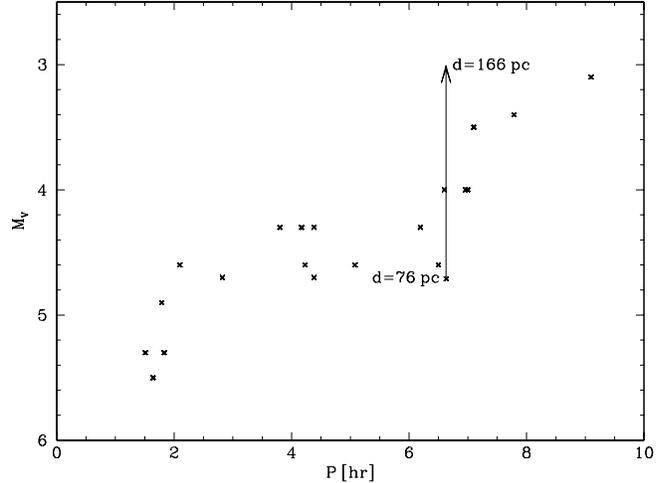}
\caption[]{The absolute magnitudes of dwarf novae adapted from
\citet{warner87-1} corrected to a mean inclination of $i=56^{\circ}.7$
and with $m_{\mathrm{v}}=8.5$ 
for the maximum
magnitude of SS\,Cyg.  Due to the increased distance, SS\,Cyg has
become the absolute brightest dwarf nova.}
\end{figure}

\subsection{The mean mass transfer rate}
Figure\,3 shows a light curve computed with the FE/FD DIM code
used for the first time in \citet{schreiber+hessman98-1} and described
in detail in \citet{schreiberetal00-1}. We adapted the parameters
suggested by \citet{ludwigetal94-1} and two different values for the
mass transfer rate. 
\begin{figure}
\label{f-fefd}
\includegraphics[angle=270,width=8.35cm]{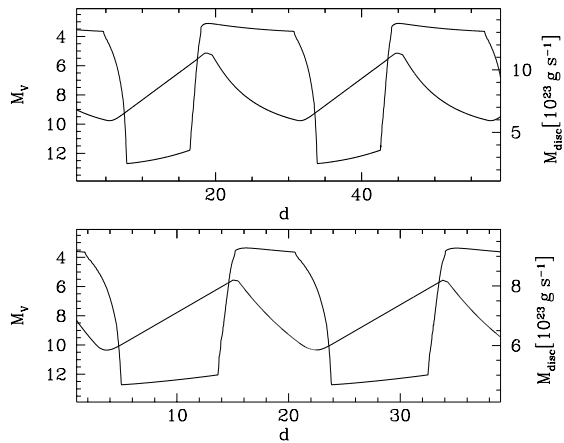}
\caption[]{Simulated light curves using the orbital parameters of SS\,Cyg,
$\alpha_{\mathrm{h}}=0.5$, $\alpha_{\mathrm{c}}=0.05$,
$\log\,(\dot{M}_{\mathrm{tr}})y=17.7$ (upper panel) and
$\log\,(\dot{M}_{\mathrm{tr}})=17.4$.}
\end{figure}
%
%In principle one should use simulated light curves, such as those
%shown in Fig.\,3, and select the parameter for which the observed
%light curve is reproduced. This happens to be impossible because
%precisely the model cannot fit the observations; it is quite clear from
%Fig.\,3 that one cannot get a system as brighter than $\sim\,3.0--3.5$. 
%  
%we want to keep this study as much as possible
%independent on the details of numerical simulations in order to check the
%validity of the current version of the DIM in general.

During an outburst cycle the disc subsequently goes through a
phase of quiescence, rise, adjustment, quasi-stationary outburst
state, and decay.  After the heating front has brought the entire disc
into the hot state (rise), the disc redistributes its mass
(adjustment) until it accretes without radial variations of the
accretion rate (quasi-stationary state).  Almost all the mass accreted
during an outburst cycle is transferred onto the white dwarf during
this quasi-stationary outburst state (Fig.\,3).  Thus, to derive a
lower limit for the mass accreted during a mean outburst cycle, one
simply has to multiply the accretion rate of a stationary standard
disc which reaches $m_{\mathrm{V}}=8.5\pm\,0.1$ with the mean
quasi-steady state duration,
$t_{\mathrm{qs}}\sim\,5.76$\,d\footnote{Assuming an adjustment time of
$t_{\mathrm{ad}}\sim\,2$\,d.}.  Throughout this paper we assume a very
large disc for SS\,Cyg during outburst, i.e. $(90\pm\,5)$\% of the
primary's Roche radius $R_{L1}$. 
\citep{ritter80-1,harrop-allin+warner96-1}
($0.9\,R_{L1}=5.83\times\,10^{10}$ cm for the binary parameter of
SS\,Cyg).  Following \citet{buat-menardetal01-1}, we describe the
additional heating of the outer disc due to the stream disc impact
with
\begin{equation}
Q_{\mathrm{i}}=\eta_{\mathrm{i}}\frac{G\,\Mwd\,\dot{M}_{\mathrm{tr}}}{2\,R_{\mathrm{d}}}\frac{1}{2\pi\,R_{\mathrm{d}}\,\Delta\,R_{\mathrm{hs}}}\exp{\left(\frac{R_{\mathrm{d}}-R}{\Delta\,R_{\mathrm{hs}}}\right)},
\end{equation}
where $R_{\mathrm{d}}$ is the outer disc radius and
$\eta_{\mathrm{i}}$ the efficiency of disc heating.  In this section
we assume $\eta_{\mathrm{i}}=1.0\pm\,0.25$ and the Keplerian kinetic
energy being released over a radial extent of
$\Delta\,R_{\mathrm{hs}}=0.1\,R_{\mathrm{d}}$ with an exponential
attenuation \citep{buat-menardetal01-1}. In addition, we include
irradiation of the disc by the white dwarf and the boundary layer
using Eq.\,(4) and (5) of \citet{schreiberetal00-1} with an albedo of
the accretion disc of $\beta=0.75\pm\,0.25$. To account for the
contribution of the irradiated secondary (K4/5) we used the code
BINARY++ written by A. van Teeseling \citep{vanteeselingetal98-1} and
found even for an extremely high accretion rate of
$\dot{M}_{\mathrm{out}}=4\times\,10^{18}$g\,s$^{-1}$ the visual
magnitude of the irradiated secondary being
$m_{\mathrm{v,sec}}\geq\,11$. We consider $m_{\mathrm{v,sec}}=\,11$ as
an upper limit for the contribution of the irradiated secondary
throughout  this paper.
\begin{table}
\caption{Assumed system parameter}
\begin{tabular}{l@{\,=\,}r@{~~}l@{\,=\,}r}
\hline\noalign{\smallskip}
\multicolumn{2}{c}{Binary parameter} & \multicolumn{2}{c}{Outburst parameter} \\
\noalign{\smallskip}\hline\noalign{\smallskip}
$\Mwd$ & $1.19\pm0.05$ $M_{\odot}$       & $t_\mathrm{out}$ & 10.76\,d\\
$M_{\mathrm{sec}}$ & 0.70 $M_{\odot}$    & $t_\mathrm{qui}$ & 38.71\,d\\
$i$   & $37^{\circ}\pm5^{\circ}$         & $t_\mathrm{cyc}$ & 49.47\,d\\
$P$   & 6.6\,hr                          & $t_\mathrm{ris}$ & 0.5\,d  \\
$d$   & $166\pm12$\,pc                   & $t_\mathrm{dec}$ & 2.5\,d  \\
$R_\mathrm{d}$ & $0.90\pm0.05R_{L1}$     & $m_\mathrm{V}$   & $8.5\pm0.1$\\
%      & $(5.83\pm0.33)\times10^{10}$\,cm  \\
%
\noalign{\smallskip}\hline\noalign{\smallskip}
\multicolumn{4}{c}{Stream impact and irradiation heating efficiencies}\\
\multicolumn{2}{c}{$\eta_i=1.0\pm0.25$}   & \multicolumn{2}{c}{$(1-\beta)=0.25\pm0.25$}\\
\noalign{\smallskip}\hline\noalign{\smallskip}
\multicolumn{4}{c}{Contribution of the irradiated secondary}\\
\multicolumn{4}{c}{$m_\mathrm{V,sec}\ge\,11$}\\
\noalign{\smallskip}\hline\noalign{\smallskip}
\end{tabular}
\end{table}

The accretion rate necessary to reproduce the observed visual
flux during the outburst can be described with a power-law:
\begin{eqnarray}
& &\dot{M}_{\mathrm{out}}=3.62\times\,10^{18}\,\mathrm{g\,s}^{-1}\left(\frac{d}{166\mathrm{pc}}\right)^{3.990}\left(\frac{i}{37^{\circ}}\right)^{1.499}
\nonumber\\& &\hspace{0.8cm}
\times
\left(\frac{\Mwd}{1.19\,M_{\odot}}\right)^{-1.449}
\left(\frac{\eta_{\mathrm{i}}}{1.0}\right)^{-0.023}
\left(\frac{R_{\mathrm{10}}}{5.83}\right)^{-1.130}
%\left(\frac{R_{\mathrm{d}}/R_{L1}}{0.9}\right)^{-1.130}
\nonumber\\& &
\hspace{1.2cm}
\times
\left(\frac{\beta}{0.75}\right)^{0.058}
\left(\frac{m_{\mathrm{V}}}{8.5}\right)^{-15.508}.
\end{eqnarray}
With the parameters and their uncertainties given in Table\,1., this
results in
%We obtain an accretion rate during outburst of
%
%%%%%%%%%%%%%%%%%%%%%%%%% fuer inclination nur mit cos(i)%%%%%%%%% 
%                                                                %
%  $\dot{M}_{\mathrm{out}}=2.81\times\,10^{18}$\,g\,s$^{-1}$.    %
%                                                                %
%%%%%%%%%%%%%%%%%%%%%%%%%%%%%%%%%%%%%%%%%%%%%%%%%%%%%%%%%%%%%%%%%%
%
%%%%%%%%%%%   mit limb-darkening (Eq. 1) %%%%%%%%%%%%%%%%%%%%%%%%%
\begin{equation}
\dot{M}_{\mathrm{out}}=(3.62\pm\,1.47)\times\,10^{18}$\,g\,s$^{-1}. 
\end{equation}
The accreted mass has to be transferred to the disc during a mean outburst
cycle by a  mean mass transfer rate:
%%%%%%%%%%%%%%%%%% nur cos (i) %%%%%%%%%%%%%%%%%%%%%%%%%%%%%%%%%%%%
%                                                                 % 
%\begin{equation}
%\overline{\dot{M}}_{\mathrm{tr}}\gappr\frac{\dot{M}_{\mathrm{out}}\,5.76}{49.47}=3.27\times\,10^{17}\mathrm{g\,s}^{-1}.
%\end{equation}                                                  
%                                                                  %
%%%%%%%%%%%%%%%%%%%% ohne stream impact %%%%%%%%%%%%%%%%%%%%%%%%%%%%
%\begin{equation}
%\overline{\dot{M}}_{\mathrm{tr}}=\frac{\dot{M}_{\mathrm{out}}\,5.76}{49.47}=3.92\times\,10^{17}\mathrm{g\,s}^{-1}.
%\end{equation}
%
%%%%%%%%%%%%%%%%%%%% mit strom und Eq.(1) %%%%%%%%%%%%%%%%%%%%%%%%%%
\begin{equation}
\overline{\dot{M}}_{\mathrm{tr}}\sim\frac{\dot{M}_{\mathrm{out}}\,t_{\mathrm{qs}}}{t_{\mathrm{cyc}}}=(4.2\pm\,1.7)\times\,10^{17}\mathrm{g\,s}^{-1}.
\end{equation}                                                  
This value has to be compared with the critical mass transfer rate.

\subsection{The critical mass transfer rate}
In the absence of stream impact heating the critical accretion rate is given
by the accretion rate at the upper turning point of the
S-curve $\dot{M}_{\mathrm{A}}$ and
can be approximated by a power-law, i.e.
\begin{equation}
\dot{M}_{\mathrm{A}}=\dot{M}_{\mathrm{crit}}=9.5\times\,10^{15}\,\mathrm{g\,s^{-1}}\,R_{10}^{2.64}\,\Mwd^{-0.88},
\end{equation}
where $R_{10}=R/10^{10}$\,cm 
\citep[see
e.g.][]{schreiber+gaensicke01-1,hameuryetal98-1,ludwigetal94-1}\footnote{The
weak dependence of $\dot{M}_{\mathrm{A}}$ on $\alpha_{\mathrm{h}}$ obtained
by \citet{hameuryetal98-1} and \citet{schreiber+gaensicke01-1} is negligible
in the context of this paper.}.

Taking into account heating of the outer edge of the disc by the
impact of the stream, the critical mass transfer rate is given by an
implicit equation as the turning point of the modified S-curve
$\dot{M}_{\mathrm{A,i}}$ depends on the mass transfer rate
itself;
$\dot{M}_{\mathrm{A,i}}=\dot{M}_{\mathrm{A,i}}(R_{10},\,\Mwd,\,\dot{M}_{\mathrm{tr}})$.
The critical mass transfer rate is then given by a solution of
\begin{equation}
\dot{M}_{\mathrm{tr}}=\dot{M}_{\mathrm{A,i}}(R_{10},\,\Mwd,\,\dot{M}_{\mathrm{tr}}),
\end{equation}
\citep[see also][]{stehleetal01-1}. Note that because the stream impact
heating decreases with decreasing radius, the critical mass transfer
rate is not necessarily defined by the vertical structure at the outer
edge of the disc.  In order to solve Eq.\,(7), we modify the
approximation of the upper turning point of the S-curve by including
impact heating. We obtain for the parameter range of SS\,Cyg:
\begin{equation}
\log(\dot{M}_{\mathrm{A,i}})=\log(\dot{M}_{\mathrm{A}})-0.039\frac{Q_{\mathrm{i}}}{10^{10}\mathrm{erg\,s}^{-1}}.
\end{equation}
Using this description, we solve Eq.(6) and obtain a series of solutions.  The
results can be approximated with the 
following power law:  
\begin{eqnarray}
& &\dot{M}_{\mathrm{crit,i}}=3.18\times\,10^{17}\,{\mathrm{g\,s}}^{-1}
%\left(\frac{R_{\mathrm{d}}/R_{L1}}{0.9}\right)^{2.750}
\left(\frac{R_{\mathrm{10}}}{5.83}\right)^{2.750}
\nonumber\\& &
\hspace{2cm}
\times\left(\frac{\eta_{\mathrm{i}}}{1.0}\right)^{-0.267}
\left(\frac{M_{\mathrm{1}}}{1.19\,M_{\odot}}\right)^{-0.287}.
\end{eqnarray}
This leads to
\begin{equation}
\dot{M}_{\mathrm{crit,i}}=(3.2\pm\,0.6)\times\,10^{17}\mathrm{g\,s}^{-1}. 
\end{equation} 

\begin{figure}
\label{f-scurves}
\includegraphics[angle=270,width=8.65cm]{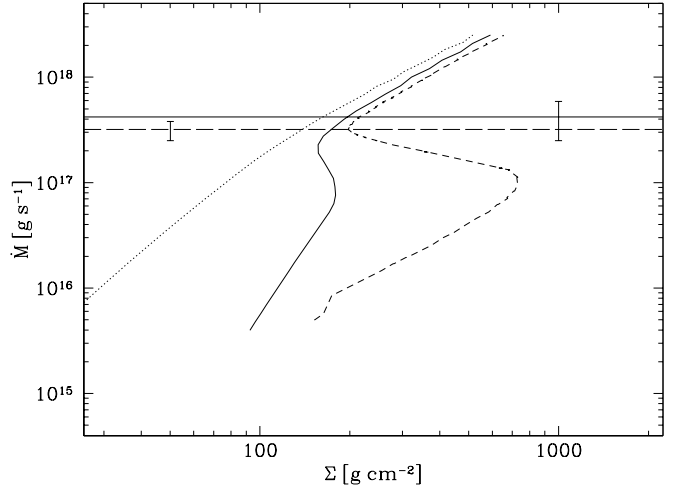}
\caption[]{S-curves for outer radii [in units of
$10^{10}$\,cm, $R=0.9\,R_{L1}=5.83$ (dotted), $0.8\,R_{L1}=5.18$
(solid), $0.7\,R_{L1}=4.53$ (short dashed line)] using
$\overline{\dot{M}}_{\mathrm{tr}}=4.2\times\,10^{17}$\,g\,s$^{-1}$
(solid horizontal line, see Eq.(5)).  At the outer edge of the disc
(dotted line) the instability is completely suppressed by the
additional heating due to the stream ($\eta_{\mathrm{i}}=1.0$). Moving
inward, the additional heating decreases (see Eq.\,(2)) but the mean
mass transfer rate (solid horizontal line) is still clearly above the
value critical for stability. The critical accretion rate
$\dot{M}_{\mathrm{crit,i}}$ (see Eq.\,(10)) is
represented by the dashed horizontal line. Notice, the S-curves are
calculated with
$\dot{M}_{\mathrm{tr}}=\overline{\dot{M}}_{\mathrm{tr}}\neq\,\dot{M}_{\mathrm{crit,i}}$.}
%The dashed-dotted line is calculated with
%$\dot{M}_{\mathrm{tr}}=\dot{M}_{\mathrm{crit,i}}=3.2\times\,10^{17}$\,g\,s$^{-1}$
%at the relevant radius $R=4.63\times\,10^{10}$\,cm.
%As stream impact heating
%does not significantly change the S-curves at smaller radii the critical
%accretion rate given by the short dashed line is quite close to
%$\dot{M}_{\mathrm{crit,i}}$.}
\end{figure}

Fig.\,4 shows S-curves for the outer edge of the
accretion disc assuming the above obtained mean mass transfer rate and
considering the additional heating by the stream (Eq.\,(2)).  The
transition between $\alpha_{\mathrm{hot}}=0.1$ and
$\alpha_{\mathrm{cold}}=0.02$ is assumed to be given by the following
temperature dependence of $\alpha$:
\begin{equation}
\begin{array}{l}
\log(\alpha)=\log(\alpha_{\mathrm{cold}})+\big[(\log(\alpha_{\mathrm{hot}})-\log(\alpha_{\mathrm{cold}})\big]\\
\hspace{3.5cm}
\times\left[1+\left(\displaystyle{\frac{2.5\times\,10^4\,\mathrm{K}}{T_{\mathrm{c}}}}\right)^8\right]^{-1},
\end{array}
\end{equation}
\citep[see][]{hameuryetal98-1}.  

For the mean mass transfer rate we obtain
$\overline{\dot{M}}_{\mathrm{tr}}\sim\,2\times\,\dot{M}_{\mathrm{A,i}}$
(Fig.\,4), which would lead to steady disc accretion. Considering the uncertainties of the involved
parameters, we find
$\overline{\dot{M}}_{\mathrm{tr}}\gappr\dot{M}_{\mathrm{crit,i}}$.

\section{The onset of the cooling front}
The critical and the mean mass transfer rate derived in the previous
section are sensitive to the uncertainties in the binary
parameters and the assumed efficiency of stream impact heating
$\eta_{i}$ as the additional heating can significantly reduce the
critical mass transfer rate.
Thus, while our results
presented in Sect.\,3 suggest that the DIM fails in describing the
observations of SS\,Cyg, they do not allow to  categorically
exclude the validity of the DIM.

However, in this section we show in an independent way that the observed light
curve of SS\,Cyg provides additional evidence for the failure of the DIM, 
independent from the details of stream impact heating or the uncertainties
in the system parameters. 
For the mass accretion rate at the onset of the decline
$\dot{M}_{\mathrm{otd}}$ the DIM
predicts 
an essentially lower value than required by the 
observations as, {\bf (a)} the DIM predicts that the start of the cooling
front at the outer edge 
of the disc defines the onset of the decline in the outburst light curve.  
The cooling front starts when the 
accretion rate of the quasi stationary disc is exactly equal to the
critical rate, thus,
\begin{equation}
\dot{M}_{\mathrm{otd,DIM}}=\dot{M}_{\mathrm{crit,i}}=(3.2\pm\,0.6)\times\,10^{17};
\end{equation}
{\bf (b)} the light curve of SS\,Cyg
shows that there is no noticeable difference between the absolute magnitude at
the onset of the decline and the outburst magnitude $m_v=8.5\pm\,0.1$
(Fig.\,1) and therefore, the observed light curve constrains
the accretion rate at this point to be equal to the outburst accretion rate
\begin{equation}
\dot{M}_{\mathrm{otd,obs}}\sim\dot{M}_{\mathrm{out}}=(3.62\pm\,1.47)\times\,10^{18}$\,g\,s$^{-1}.
\end{equation} 
The obtained huge discrepancy between the prediction of the DIM for
$\dot{M}_{\mathrm{otd}}$ (Eq.\,(12)) and the value required to explain the
observed visual magnitude (Eq.\,(13)) clearly confirms the failure of the DIM.

It is worth noting that even if we completely neglect stream impact heating 
the DIM cannot explain the observations. Then, the critical mass transfer rate
can be approximated by 
Eq.\,(6) and the accretion rate necessary to reach the visual flux 
during outburst somewhat increases:
\begin{eqnarray}
\dot{M}_{\mathrm{otd,DIM}}&=&\dot{M}_{\mathrm{crit}}=(7.6\pm\,2.0)\times\,10^{17}\mathrm{g\,s}^{-1},\\
\dot{M}_{\mathrm{otd,obs}}&\sim\,&\dot{M}_{\mathrm{out}}=(3.71\pm\,1.51)\times\,10^{18}\mathrm{g\,s}^{-1},\\
\overline{\dot{M}}_{\mathrm{tr}}&=&(4.3\pm\,1.8)
\times\,10^{17}\mathrm{g\,s}^{-1}.  
\end{eqnarray}
Obviously, the  problem for the DIM obtained in Sect.\,3 vanishes
if we neglect stream impact heating of the outer disc. Nevertheless, the DIM
clearly fails as $\dot{M}_{\mathrm{otd,obs}}\gg\,\dot{M}_{\mathrm{otd,DIM}}$.

\section{Discussion and Conclusion}
Considering the ``larger'' distance of SS\,Cyg and the observed long
term light curve, we calculated the mean mass transfer rate
$\overline{\dot{M}}_{\mathrm{tr}}$. 
Taking into account stream
impact heating of the outer disc we find
$\overline{\dot{M_{\mathrm{tr}}}}\sim\,2\times\dot{M}_{\mathrm{crit}}$
(Fig.\,4) which contrasts with the predictions of the DIM: with such a
high mean mass transfer rate, SS\,Cyg should not be a dwarf nova at
all, but a novalike variable with a stationary hot accretion disc.
This result is, however, sensitive to the assumed efficiencies of
stream heating and disk irradiation 
and the uncertainties of the binary parameters, the
inclination, and the distance. 
While this result on its own is not sufficient to proclaim a
general failure of the DIM, an independent suggestion for a major
problem with the DIM comes from comparing the accretion rate predicted
by the DIM at the onset of the decline with the accretion rate
required by the observed visual magnitude at this point (see
Sect.\,4).  Hence, the conclusion from both these results must
be: the current DIM cannot explain the observations of SS\,Cyg for
a distance of $166\pm\,12$\,pc.

There are three possibilities that may resolve this problem:
(a) the temperature dependence of $\alpha$, the key ingredient in the
accretion disc limit cycle model, has to be modified, (b) increased
mass transfer during the outburst plays an important r\^ole, or
finally (c) the HST/FGS distance of 166\,pc and/or the assumed
inclination are wrong.

Considering (a), it seems that we do not yet have an adequate
understanding of the viscosity in accretion discs.  The instability
necessary to obtain the limit cycle behaviour appears mainly due to an
artificial change of $\alpha$, given by Eq.\,(11), at the point where
the small ``natural'', i.e. without a change of $\alpha$, ionisation
instability has been found in the S-curves.  There is no physical
reason for this assumption and it might be the case that $\alpha$
starts to change at higher temperatures than previously assumed.
This would lead to a higher value of the critical mass transfer
rate. As such a modification would affect the dividing line between
dwarf novae and nova-likes, one should carefully consider the
consequences for other dwarf nova before changing Eq.\,(11).

With respect to (b), we point out that another possibility to solve
the problem obtained in Sect.\,2\,--\,4 is to assume that the mass transfer
increases during the outbursts, e.g. due to irradiation of the
secondary star.  Assuming a mean mass transfer rate of
$\overline{\dot{M}}_{\mathrm{tr,qui}}=1.5\times\,10^{17}$\,g\,s$^{-1}$
during quiescence, rise, and decay, the mean mass transfer rate during
the outburst state should be
\begin{eqnarray}
\overline{\dot{M}}_{\mathrm{tr,out}}\,&\sim&\,\frac{
\displaystyle\dot{M}_{\mathrm{out}}\,t_{\mathrm{qs}}-(t_{\mathrm{qui}}+t_{\mathrm{ris}}+t_{\mathrm{dec}})\overline{\dot{M}}_{\mathrm{tr,qui}}}
{\displaystyle t_{\mathrm{out}}-t_{\mathrm{ris}}-t_{\mathrm{dec}}}
\nonumber\\
&=&2.4\times\,10^{18}\mathrm{g\,s}^{-1},
\end{eqnarray}
i.e. increased over the mean quiescent value by a factor of $\gappr\,15$.
In this scenario the onset of the decline in the light curve would represent 
decreasing mass transfer and not the start of the cooling front. 

Finally, considering (c), we can only note that at least one of the
additional CV parallaxes measured with HST, namely U\,Gem, agrees very
well with previous distance estimates \citep{harrisonetal99-1}.
U\,Gem has a noticeably lower accretion rate than SS\,Cyg, therefore
the distance estimated based on $K$ band magnitude and/or spectroscopy
of the secondary star is much less affected by emission from the disc
than in SS\,Cyg. Thus, at least for U\,Gem an error in the HST/FGS
parallax appears extremely unlikely.
However, assuming $i=37^{\circ}$ and demanding 
$\dot{M}_{\mathrm{out}}\,\sim\,1.0\times\,10^{18}$\,g\,s$^{-1}$ for SS\,Cyg 
leads to
$\overline{\dot{M}}_{\mathrm{tr}}\,\sim\,1.2\times\,10^{17}$\,g\,s$^{-1}$
and requires $d\,\sim\,117$\,pc for agreement with the DIM. 

\begin{acknowledgement}
This research has made use of the AFOEV database, operated at CDS,
France.  Special thanks are extended to Jean-Marie Hameury, Hans Ritter and  
Rick Hessman for interesting discussions at the
CV-Conference in G\"ottingen (2001). We also wish to thank an anonymous
referee for helpful comments. MRS thanks the Astronomy
Department of the UCT for the warm hospitality during a nice and
productive stay in March\,--\,May 2001 and the DAAD (PKZ:D/01/05718)
for financial support.  BTG thanks for support from the DLR under
grant 50\,OR\,99\,03\,6.
\end{acknowledgement}

\bibliographystyle{apj}

%\bibliography{../../../../aamnem99,../../../../aabib}
%\bibliographystyle{../../../../aacite}
%\bibliography{aamnem99,/home/cthulhu/boris/tex/Papers/Bibliography/aabib}

%$t_{\mathrm{cyc}}=49.47$d,
%$t_{\mathrm{out}}=10.76$d,
%$t_{\mathrm{ris}}=0.5$d
%$t_{\mathrm{dec}}=2.5$d
%$t_{\mathrm{qui}}=38.71$d. 
\end{document}